\documentclass[%
 reprint,
superscriptaddress,
nofootinbib,
 amsmath,amssymb,
 aps,
 onecolumn,
 notitlepage
]{revtex4-1}

\usepackage{graphicx}
\usepackage{dcolumn}
\usepackage{bm}

\usepackage{braket}     
\usepackage{xcolor}     
\usepackage{enumitem}   
\usepackage{listings}   

\def\Tr{\text{Tr}}

\begin{document}

\title{QAOA for Max-Cut requires hundreds of qubits for quantum speed-up}

\author{G.G. Guerreschi}
 \email{gian.giacomo.guerreschi@intel.com}
 \affiliation{Intel Labs, Intel Corporation, Santa Clara, CA 95054, USA}
\author{A.Y. Matsuura}
 \affiliation{Intel Labs, Intel Corporation, Santa Clara, CA 95054, USA}
\date{\today}


\begin{abstract}
Computational quantum technologies are entering a new phase in which noisy intermediate-scale quantum computers are available, but are still too small to benefit from active error correction. Even with a finite coherence budget to invest in quantum information processing, noisy devices with about 50 qubits are expected to experimentally demonstrate quantum supremacy in the next few years. Defined in terms of artificial tasks, current proposals for quantum supremacy, even if successful, will not help to provide solutions to practical problems.
Instead, we believe that future users of quantum computers are interested in actual applications and that noisy quantum devices may still provide value by approximately solving hard combinatorial problems via hybrid classical-quantum algorithms. To lower bound the size of quantum computers with practical utility, we perform realistic simulations of the Quantum Approximate Optimization Algorithm and conclude that quantum speedup will not be attainable, at least for a representative combinatorial problem, until several hundreds of qubits are available.
\end{abstract}

\pacs{Valid PACS appear here}
\maketitle



Quantum computers are soon going to satisfy the requirements, both in terms of number and quality of qubits, to perform restricted forms of computation that are beyond what is classically feasible in a reasonable time. This milestone is often called quantum supremacy \cite{Preskill2012} and is expected to be achieved in the next couple of years with systems of about 50 qubits. While impressive, these demonstrations will not bring practical advantage to potential users of quantum computers. For example, the random quantum circuits sampling problem that may be used to show quantum supremacy \cite{Boixo2016,Neill2018} will not help us solve problems of practical interest.

For users interested in solving real-world applications, it is natural to use quantum speedup to quantify the advantage of quantum computers over classical machines. Since quantum algorithms have a computational cost that is either polynomially \cite{Grover1997} or exponentially lower \cite{Shor1999} than their classical counterparts, for larger problem instances quantum algorithms provide an increasing advantage over classical approaches.

However, the quest for quantum speedup has to adapt to the current state of the technology, in which devices have a relatively small number of qubits, all of limited quality. We are entering the NISQ era, the era of Noisy Intermediate Scale Quantum devices \cite{Preskill2018}. Without active error correction, with a limited coherence budget and with realistic constraints in qubit connectivity, it is questionable whether we can expect any speedup at all and unclear what size of quantum computers one would need to attain speedup.

In advance of mature quantum computing systems able to address these questions experimentally, powerful simulation tools are needed to reproduce the evolution of quantum systems in the presence of noise and benchmark short-term quantum algorithms. In particular, heuristic algorithms are recently attracting much attention due to their resilience to certain systematic errors. However, it is essential to understand an algorithm's behavior under realistic hardware errors, as this insight helps us determine which algorithms represent concrete alternatives to classical methods at the current stage of quantum technology.

In this Article we investigate the Quantum Approximate Optimization Algorithm (QAOA) \cite{Farhi2014}, a hybrid quantum-classical algorithm, and compare its computational cost with a state-of-the-art classical solver (AKMAXSAT \cite{Kugel2012}) for the NP-hard problem called Max-Cut. The solution of Max-Cut, even if approximate, has practical application in machine scheduling \cite{Alidaee1994}, image recognition \cite{Neven2008} or for the layout of electronic circuits \cite{DEZA1994191}. In our simulations we include both decoherence and dissipation to verify that the quantum algorithm is robust to realistic levels of noise, and consider the compilation of the algorithm in terms of one- and two qubit gates satisfying the hardware connectivity of a square grid with the corresponding routing overhead. Multiple instances at different problem sizes are solved to estimate the absolute time required to run QAOA and extrapolate its scaling cost.

Our findings indicate that solving Max-Cut, a graph partitioning problem, for small graphs with tens of nodes with a quantum computer takes the same time as solving much larger graphs with hundreds of nodes with classical solvers. Based on data from our extensive simulations at small system sizes, and therefore coping with significant extrapolation uncertainty, one can estimate that the performance crossover between the two approaches would require between several hundreds and a few thousand qubits.


\section*{Results}

\subsection*{The Quantum Approximate Optimization Algorithm}
The algorithm chosen for this study, namely the Quantum Approximate Optimization Algorithm (QAOA), belongs to the class of hybrid algorithms and requires, in addition to the execution of shallow quantum circuits, a classical optimization process to improve the quantum circuit itself. Since its proposal \cite{Farhi2014}, QAOA has attracted considerable interest as an application of pre-error-corrected quantum devices \cite{Farhi2016a,Guerreschi2017a,Farhi2017a,Wang2017a,Hadfield2017a}.

The goal of QAOA is to solve, in an approximate way and by using a shallow quantum circuit, combinatorial problems like constraint-satisfaction (SAT) problems in their ``maximization'' formulation. In general, SAT problems are characterized by a set of constraints and by the question of whether it is possible to satisfy all constraints at the same time or not. Their ``max'' form asks what is the maximum number of constraints that can be satisfied at the same time. The latter question is at least as hard as the former one.

What differentiates it from other variational algorithms is that the quantum circuit is determined by the problem instance with very few adjustable parameters, and it is believed that shallow circuits suffice to achieve good approximate solutions.
We apply QAOA to solve Max-Cut, a graph partitioning problem defined as follows. Given a graph, assign one of two colors (e.g., black or white) to each vertex. An edge can be ``cut'' when it connects two vertices of different color. The problem is to assign the colors so that as many edges as possible can be cut at the same time. Max-Cut belongs to the complexity class called NP-hard, and closely related problems have been used as benchmarks in the context of Adiabatic Quantum Optimization and, in particular, with D-Wave Inc. devices \cite{Boixo2014}.
The objective function corresponds to the hermitian operator:
\begin{equation}
    C = -\frac{1}{2} \sum_{(i,j) \in E} Z_i Z_j
\end{equation}
where $E$ is the set of edges of the graph to partition, $Z_i$ the Z Pauli matrix of (logical) qubit $i$, and we have neglected an overall constant.
The QAOA circuit is obtained by repeating two blocks of quantum operation for a total of $p$ times, with $p$ being an adjustable parameter. The variational state prepared by the QAOA circuit has the form:
\begin{equation}
    \ket{\gamma,\beta} = e^{-i \beta_p B} e^{-i \gamma_p C} \dots e^{-i \beta_1 B} e^{-i\gamma_1 C}\ket{+\ldots+}
\end{equation}
with $B=\sum_k X_k$ being the sum of the X Pauli matrices associated to each vertex of the graph. Notice that the depth of the quantum circuit is approximately linear in the parameter $p$. Expectations are that the number of parameters (given by $2p$) and circuit depth scale lightly with the problem size.

Notice that the complexity of Max-Cut depends on the regularity and connectivity of the graph itself. Following an existing trend in the community \cite{Farhi2014,Guerreschi2017a,Farhi2017a,Hadfield2017a}, we consider random 3-regular graphs, defined as having exactly 3 edges per vertex. Extending the analysis to other types of graph is relatively simple, but the numerical results may differ.


\subsection*{Compiling and scheduling the quantum circuits}
The quantum circuit to prepare state $\ket{\gamma,\beta}$ is intuitively expressed in terms of single qubit rotations for the $B$ part and two-qubit phase gates for the part involving $C$. Rotation angles and phases are free to vary in a continuous way in QAOA and their implementation with a finite set of gates (think for example of the universal set $\{H,T,CNOT\}$) may be daunting. However, it is easy to envision that the variational parameter is effectively the time-integrated amplitude of the control field acting on the qubits during the rotation: amplifying the signal amplitude corresponds to increasing the rotation angle. Similar considerations are valid for the phase gates, either implemented directly via controlled $ZZ$ interactions (tuning the coupling strength or pulse scheme) or in terms of two CNOTs and a rotation (in this case all two-qubit gates are independent of the variational parameters).

Due to limited hardware connectivity, an additional compilation step is necessary to produce executable circuits. A valid schedule must satisfy three constraints: 1) the logical dependencies between the gates forming the algorithm, 2) the exclusive activation of a qubit, which can be involved in at most one gate at a time, and 3) the fact that two-qubit gates can be performed only between connected qubits \cite{Venturelli2018,Guerreschi2018}. Here we consider a square grid connectivity and use a novel scheduler approach \cite{Guerreschi2018} that introduces SWAP operations to route the qubits while minimizing the overall circuit depth. For simplicity, we consider that all gates (single-qubit rotations forming the B operator, controlled-phase gates forming C, and SWAP gates for routing) have the same duration. The circuit depth is the relevant metric, even more than the number of gates, when the main source of error is due to decoherence and not to imperfect pulse control. In Appendix~\ref{app:scheduling}, we provide a full schedule for a typical 8-qubit instance of QAOA with $p=4$ and show a visual representation of the first few parallel gates.


\begin{figure}[b]
\vspace{-1mm}
\includegraphics[scale=0.5]{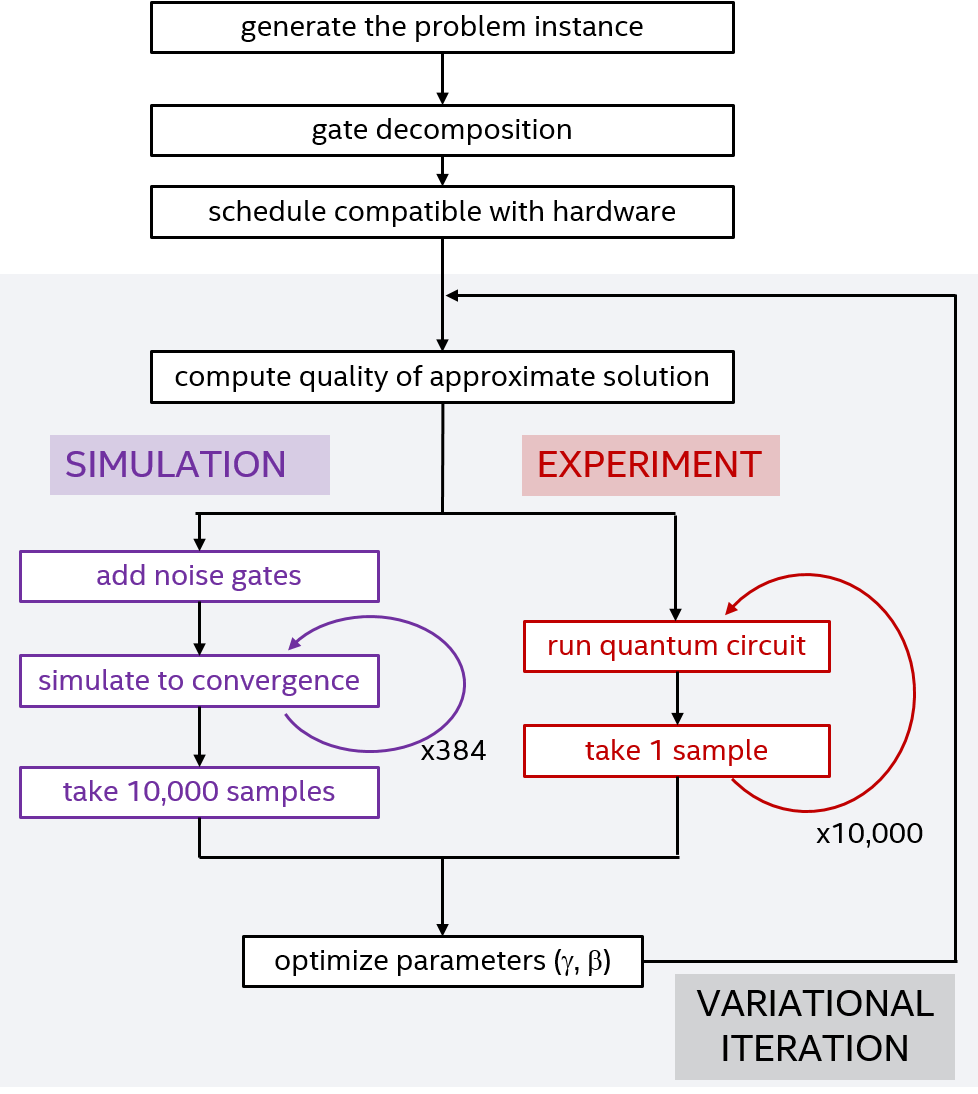}
\centering
\caption{Schematic diagram of the QAOA algorithm and auxiliary tasks to solve a single Max-Cut instance. The two branches, labelled ``simulation'' and ``experiment'' respectively, distinguish between the operations to simulate the variational algorithm with classical computers and those to perform it experimentally with quantum devices. In our study, 10000 samples are used to estimate the value of the cost function $\bra{\gamma,\beta}C\ket{\gamma,\beta}$ at each variational iteration.}
\label{fig:diagram}
\vspace{-3mm}
\end{figure}

\begin{figure*}[th]
\vspace{-1mm}
\includegraphics[scale=0.56]{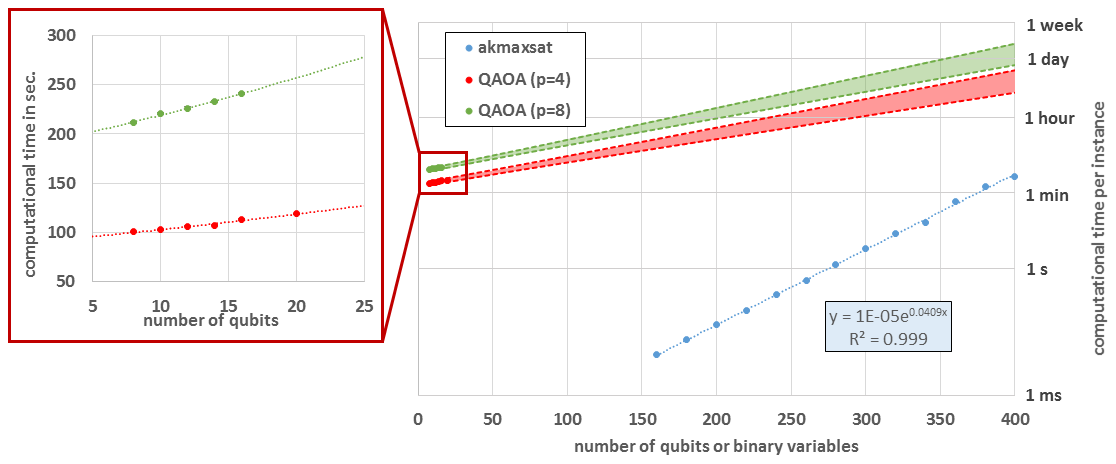}
\centering
\caption{\textbf{Main panel: } Computational cost of solving a single Max-Cut instance on random 3-regular graphs. Blue markers correspond to the classical baseline (AKMAXSAT solver) while red and green marks correspond to the experimental time required by the quantum algorithm QAOA, with $p=4$ and $p=8$ respectively. The error bars for the single data points are smaller than the markers (see Appendix~\ref{app:QAOA-statistics}). Notice that in the time needed by QAOA to partition graphs with 20 vertices, AKMAXSAT partitions graphs about 20 times larger.
The blue dashed line is the result of a fitting procedure with an exponential function. The red and green areas are associated with a 95\% confidence interval for the prediction of the QAOA cost based on a linear regression of $\log_{10}(T)$ as a function of the number of qubits (here $T$ is the computational time per instance). This extrapolation should be seen as suggesting a qualitative behavior due to the uncertainty in the extrapolation from relatively small system sizes.
\textbf{Insert panel: } Magnification of QAOA datapoints. Notice that exponential curves, and smooth curves in general, locally resemble straight lines and this makes it difficult to exclude other functional forms for the extrapolation. It is, however, believed that even quantum computers will not be able to solve NP-hard problems in polynomial time.}
\label{fig:main-result}
\vspace{-2mm}
\end{figure*}


\subsection*{Quantum simulations with realistic noise}
The simulation of a quantum system is a computationally intensive task, requiring an amount of memory and number of operations that both scale exponentially in the number of qubits. The massively parallel simulator named qHiPSTER (Quantum HIgh Performance Software Testing EnviRonment) has recently been developed to take maximum advantage of multi-processor and multi-node architectures for the simulation of ideal quantum circuits \cite{Smelyanskiy2016a,Thomas2016}. We adapted it by introducing single-qubit decoherence and dissipation using the Noise Gate approach \cite{Bassi2008}. This method considers that noise acts on each qubit separately (no spatial nor time correlations) in between the execution of gates. By unraveling the master equation in terms of a stochastic Schr\"edinger equation, the authors of \cite{Bassi2008} obtained a mathematical equivalence between the density matrix at the end of the circuit and the average over an ensemble of pure states, each evolved via a stochastically ``perturbed'' quantum circuit. The perturbation comes from including extra noise gates, after each actual circuit operation, to emulate the effect of noise. In practice, each noise gate corresponds to a single-qubit rotation around a stochastic axis and by a stochastic angle satisfying a suitable Gaussian distribution \cite{Smelyanskiy2016a}, and each perturbed circuit is realized by choosing the rotation axis and angle for every noise gate independently.

Reference \cite{Sawaya2016} describes the first implementation of noise gates in the qHiPSTER simulator, but in that study the circuit was considered composed by sequential gates. Here we extended the implementation by taking into account the parallelism between gates, as expressed in the optimized schedule, with the effect of reducing the number of noise gates to simulate.
Notice that the simulation only deals with pure states, but now it requires to be repeated multiple times (one repetition per noise gate realization) for each quantum circuit. Therefore part of the parallelism of qHiPSTER has been devoted to the contemporaneous simulation of the perturbed circuits necessary to achieve convergence: as discussed in Appendix~\ref{app:convergence}, we choose to average over 384 noise realizations.


\subsection*{Computational time cost of QAOA}
Since the goal of our work is to understand the potential of quantum hardware, the computational time required by QAOA is not the time to simulate the quantum algorithm, but the absolute time it would have taken to implement it on actual devices. We report the time needed to solve a single instance, obtained by averaging the results of 40 instances at each problem size in the presence of noise.

Despite recent proposals of gradient-based methods, like gradient descent or Quasi-Newton methods, in the context of QAOA \cite{Guerreschi2017a} and other variational algorithms \cite{Romero2017a}, we used the typical choice of the gradient-free Nelder-Mead method \cite{Nocedal2000} due to the relatively small dimensionality of the parameter space and the high sampling cost of gradient estimation. The Nelder-Mead method is characterized by four parameters corresponding to reflection, expansion, contraction, and shrinkage (or reduction): we used the values 1.1, 1.5, 0.6, and 0.4 respectively.

Each instance requires several optimization runs that differ due to the randomized initial values of the parameters $(\gamma,\beta)$. This randomization is essential to reach the global solution since each run substantially corresponds to a local optimization, even when the Nelder-Mead simplex is initialized with no locality constraint. Furthermore, each optimization run is composed of hundreds of variational iterations (we limited each run to maximum 300 updates of the simplex, but the run can end earlier if the best simplex vertex did not change in $10p$ updates) in which the cost function $\bra{\gamma,\beta}C\ket{\gamma,\beta}$ is estimated by sampling the final state $\ket{\gamma,\beta}$ in the computational basis. As illustrated in FIG.~\ref{fig:diagram}, while this operation can be performed in negligible time during simulations, it requires one experimental repetition of the same experiment for each sample. In practice, one needs (tens of) thousands of repetitions of the same quantum circuit to have enough statistics to estimate the objective function \cite{Kandala2017,Otterbach2017a}.

The number of experimental repetitions of the coherent quantum computation to solve a single problem instance is therefore given by
(20 optimization runs)$\times$(hundreds of function evaluations per run)%
$\times$(10000 repetitions for statistics).
The time duration of a single repetition $T_\text{s.r.}$ is given by the number of sequential quantum operations per quantum circuit, and we consider it to be:
\begin{equation*}
    T_\text{s.r.} = T_P + \text{(ciruit depth)} \times T_G + T_M \, ,
\end{equation*}
with $T_P$ being the time to prepare the initial state, $T_G$ the average duration of a quantum gate, and $T_M$ the time needed to measure the qubits. Multiplying $T_\text{s.r.}$ by the number of repetitions of coherent computation provides the time cost of a single red or green marker in FIG.~\ref{fig:main-result}. The time scales used in our study reflect realistic, but aggressive, projections for state-of-the-art devices based on superconducting circuits \cite{Devoret2013,Barends2014,OMalley2016,Rol2017,Kandala2017}: $T_P+T_M=1\,\mu$s, $T_G=10\,$ns, $T_2=100\,\mu$s, and $T_1=200\,\mu$s.



\subsection*{Classical alternative}
For a meaningful benchmark, the classical baseline must be represented by a state-of-the-art solver that is able to optimize a spectrum of problems similar to the applications of QAOA. A natural choice is considering solvers for constraint-satisfaction problems (called SAT problems) and their ``max'' formulation.

A good Max-SAT solver that is openly available to the research community, and often utilized in benchmarks of Adiabatic Quantum Optimization devices \cite{Santra2014}, is represented by AKMAXSAT \cite{Kugel2012,book_akmaxsat}.
It is possible to reduce any Max-Cut problem to a Max-2-SAT problem by a simple reduction: each vertex of the graph corresponds to a distinct binary variable for the SAT, and each edge of the graph gives rise to two constraints (see Appendix~\ref{app:akmaxsat} for details).

In this case, the computational time is the average time required to solve a single instance with AKMAXSAT running on a single Intel Xeon Phi processor
(Performance results are based on testing as of December 20th 2017 and may not reflect all publicly available security updates. See configuration disclosure for details. No product can be absolutely secure. Intel, Xeon, and Intel Xeon Phi are trademarks of Intel Corporation in the United States and other countries. Other names and brands may be claimed as the property of others.).
The timings reported in FIG.~\ref{fig:main-result} and in Appendix~\ref{app:akmaxsat} are averaged over 400 instances of Max-Cut and exploited the parallelism offered by the 68 cores of the Intel Xeon Phi processor in a trivial way: each instance was separately run on a single core, effectively providing a reduction of the total computational time that is strictly linear in the number of cores. No classical strategy can provide more speedup from parallelism than a linear reduction. For consistency, the computational time of a single instance is considered to be equal to the total computational time divided by the number of instances (400 in this case).


\subsection*{Comparison and performance crossover}
The computational cost of both the classical approach and the quantum variational algorithm are showed in FIG.~\ref{fig:main-result}, the blue markers corresponding to datapoints for the classical AKMAXSAT, while the red and green markers for QAOA with $p=4$ and $p=8$ respectively. We can draw two qualitative conclusions: the first is that, in the time needed to experimentally run QAOA for 20 qubits, AKMAXSAT is able to solve Max-Cut instances corresponding to graphs with about 20 times more nodes. This means that, even in the most favorable scenario, quantum advantage cannot be achieved without hundreds of qubits. Observe also that the noise is relatively weak and corresponds to $T_2=10^4\times T_G$, i.e. 10,000 times the average gate duration (and about 100 times the measurement time $T_M$).

The second conclusion is based on the 95\% confidence interval for the predicted exponential behavior of the QAOA cost extrapolated from the small-size datapoints. It appears that, if a crossover between the quantum and classical solvers exists, it takes place for instances with several hundreds of binary variables. While for small-size instances the classical baseline is much faster, beyond the crossover point there would be opportunity for quantum speedup. However, it is important to take into consideration three additional aspects to properly interpret the latter observation.

First, the rigorous extrapolation of the behavior for large problem sizes requires having results for ``large enough'' instances to avoid being deceived by small-size effects. A priori it is unclear what large enough means in the current context, but it is computationally demanding to perform extensive simulations in the presence of noise beyond 20-24 qubits. The extrapolation is then highly influenced by the choice of the fitting function and possibly unreliable. A close-up of the quantum datasets is presented in the inset panel

Second, the quantum solver is not guaranteed to reach the best solution, but it only provides a good approximation. This adds a third relevant quantity in addition to problem size and computational time, namely the quality of the approximate solution. We verified that the quantum state produced at the end of the (optimized) QAOA circuit has substantial overlap with the global solution (see Appendix~\ref{app:gs-population}). For 20 qubits and $p=4$ the overlap is around $10\%$ and the exact solution would most probably be obtained in many of the repetitions required in the protocol. Indeed, the global solution would likely appear much earlier in the optimization process. The extension of our study to classical approximate algorithms would allow the introduction of a third axis in FIG.~\ref{fig:main-result} to characterize and compare the quality of the approximate solutions. Currently the best approximate solvers for Max-Cut are based on the reformulation of the problem in terms of semi-definite programming \cite{Goemans1995,Halperin2004} or following a combinatorial approach \cite{Bazgan2008}.

Third, it is unclear whether $p=8$ will be sufficient to solve problems involving several hundreds or a few thousands of qubits. Expectations are that $p$ should grow, even if relatively mildly, with the number of qubits $N$. While this is not a problem in terms of circuit depth (that would increase only linearly with $p$), we notice that the coefficient in the exponential function increases from 0.0141 for $p=4$ to 0.0159 for $p=8$ (the values for the upper bound of the prediction interval are 0.0155 and 0.0173 respectively, slightly larger than those for the extrapolated curves). If this observation is a trend related to increasing $p$, quantum speedup may be compromised if the coefficient overcomes the equivalent one for classical algorithms (0.0409 for AKMAXSAT) before the crossover point is reached.

Finally, a recent study \cite{Crooks2018a} analyzed QAOA for Max-Cut and reached apparently opposite conclusions supporting ``the prospects that QAOA will be an effective method for solving interesting problems on near-term quantum computers''. However the base of the comparison is very different from our study since the resource intensive part of finding the optimal (or at least good) values of the parameters is separated from the solution of the specific instance. Most of the cost of running QAOA is then absorbed into a sort of global training and neglected in the cost-per-instance analysis. While the approach is surely interesting, no specific arguments are provided to quantify the efficacy of global training for larger and larger instances and therefore their claims remain unproven.


\section*{Discussion}

In the era of noisy intermediate-scale quantum devices, the first experimental milestone within reach is the so-called Quantum Supremacy: performing certain tasks designed to be unfeasible for current and near-future classical machines.
However, potential users of quantum computers would certainly like to gain insight on practical problems either by achieving the solution faster or, in other contexts not explored in this work, by improving the quality of approximate solutions.

We performed extensive simulations of the Quantum Approximate Optimization Algorithm applied to instances of a graph partitioning problem, called Max-Cut, belonging to the NP-hard complexity class. We included all the relevant aspects to make the simulations as realistic as possible. Quantum circuits were decomposed in one- and two-qubit gates and scheduled according to the underlying qubit connectivity \cite{Guerreschi2018}, here a bidimensional square grid. Noise was introduced following an approach based on the integration of stochastic Schr\"oedinger equations \cite{Bassi2008,Sawaya2016}. The statistical uncertainty due to sampling the quantum state at the end of QAOA circuits was taken into account by simulating the very same process (using pseudo-random generators), and it impacted the performance of the classical optimizer.

Our results show that classical solvers are very competitive until several hundreds of variables are considered. Quantum speedup, in terms of the absolute time required and not in scaling, can be achieved only after the performance crossover point marking the break-even condition of quantum and classical methods. We estimate the crossover to be between several hundreds and a few thousands qubits for QAOA applied to the solution of Max-Cut instances for random 3-regular graphs.

As a final remark, it is interesting to consider how the situation changes by having at our disposal not a single, but a number of quantum devices. Following the protocol described in this work, it is trivial to parallelize the quantum computation with respect to the instance and optimization run. However, the function evaluations encountered in each optimization run must be performed sequentially if one wants to preserve the resilience of hybrid algorithms to systematic errors. In addition, one cannot use multiple quantum devices to accumulate the repetition statistics faster (in fact, each device may be affected by different calibration errors). The parallelism is then limited by the homogeneity of the quantum devices at our disposal and our desire to preserve the robustness of variational algorithms.



\begin{acknowledgments}
The authors acknowledge the Texas Advanced Computing Center (TACC) at The University of Texas at Austin for providing HPC resources that have contributed to the research results reported in this work. URL: http://www.tacc.utexas.edu
Related simulations have been performed with resources of the National Energy Research Scientific Computing Center, a DOE Office of Science User Facility supported by the Office of Science of the U.S. Department of Energy under Contract No. DE-AC02-05CH11231. The authors thank Nicolas Sawaya for the careful reading of the manuscript and his helpful feedback.
\end{acknowledgments}

\section*{Author Contributions}
G.G.G. and A.Y.M. designed the study. G.G.G. was responsible for the simulations and the numerical analysis. Both authors wrote and reviewed the manuscript.


\bibliographystyle{unsrt}
\bibliography{references}


\newpage
\onecolumngrid
\section*{\Large{Appendix}}

\vspace{2cm}

\section{Convergence of simulations with noise}
\label{app:convergence}

By design, qHiPSTER simulates noiseless quantum circuits \cite{Smelyanskiy2016a}. The way one introduces noise, following a mathematically exact approach described in \cite{Bassi2008,Sawaya2016},
is to add ``noise gates'' that cause small and stochastic perturbations of the ideal circuit. A single perturbation is not descriptive of the overall noise process, but if the simulation is
repeated with many different perturbations, then the average result captures both decoherence and relaxation.

Each perturbation requires an additional noise gate for each actual gate in the quantum algorithm, in this way doubling the total number of operations required by the simulator.
However, the main overhead comes from the need to average over the computation of a large number of different perturbations because the mathematical equivalence is only valid in that limit.
Our preliminary study indicates that several hundreds of distinct perturbations were necessary to achieve convergence of the quantity involved in the classical optimization,
at least for the noise level representative of realistic devices.

\begin{figure}[b]
\includegraphics[scale=0.7]{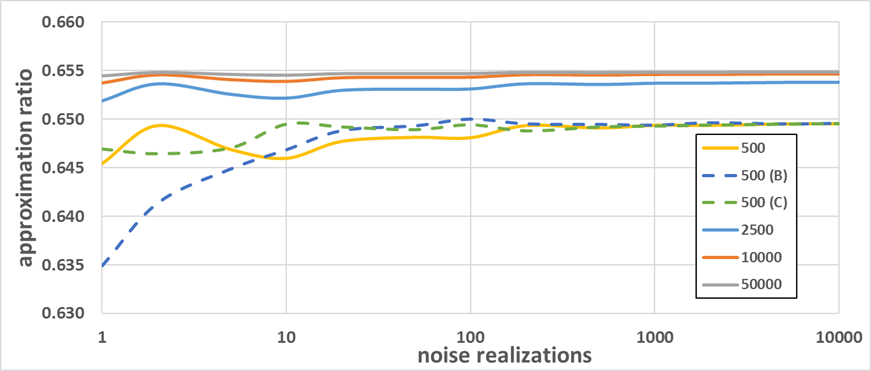}
\centering
\caption{Example of convergence for noisy simulation of 12-qubit QAOA circuits. A large number of independent perturbations of the ideal quantum circuit are simulated to reproduce the effect of noise.
Different curves relates to different noise levels as labelled by the $T_2/T_G$ ratio (with $T_2$ being the dephasing time, $T_G$ the duration of an average gate, and we considered the relaxation time $T_1=2 T_2$).
The dashed lines represent alternative realizations obtained by initializing the random number generator with different seeds.}
\label{fig:noise-convergence}
\end{figure}

FIG.~\ref{fig:noise-convergence} shows how the \emph{estimate} of the approximation ratio changes with the number of noise realizations. Considering a single line, convergence cannot be achieved
until the final plateau is reached, whose value corresponds to the asymptotic value for very large number of noise realizations. However this is a necessary but not sufficient condition,
since the number of realizations to reach the final plateau may be affected by a particularly lucky or unlucky sequence of the (random) noise gates. One needs to confirm that convergence
to the final plateau is reached irrespective of the random number stream. The dashed lines represent different noise realizations of the yellow curve at $T_G/T_2=500$ and confirm that the same
asymptotic value is reached starting from approximately 400 noise realizations.

It is important to specify how the noise level is computed. The relevant quantity is the ratio between the coherence time of the qubits and the duration of a quantum operation. While different
operations may take a different amount of time, for simplicity we consider a reasonable value of the average gate duration $T_G$. We express $T_2$, i.e. the single qubit dephasing time, as a
multiple of the gate duration. For superconducting architectures, aggressive but realistic values are: $T_2=100\,\mu$s, $T_1=200\,\mu$s and $T_G=10\,$ns \cite{Devoret2013,Barends2014,OMalley2016,Rol2017,Kandala2017}.
The corresponding curve in FIG.~\ref{fig:noise-convergence} is labelled by ``10000'', this number representing the ratio $T_2/T_G$.
In our noise model, we assume that the relaxation time $T_1$ is twice as long as the dephasing time $T_2$. This happens when the coherence is not limited by relaxation, but effectively describes
additional noise sources as a conservative expectation in short term devices.

Finally, let us compare our approach to simulate noise with alternative methods and, in particular, with the density-matrix simulators. There are two main differences: Density-matrix simulators
directly describe the evolution of the system without requiring multiple noise realizations, but a single computation is much more costly. In fact, while qHiPSTER requires the storage of $2^N$
amplitudes and $\mathcal{O}(2^N)$ operations per quantum gate, density-matrices are represented by $2^{2N}$ complex numbers and require at least $\mathcal{O}(2^{2N})$ for each update. In practice,
simulating 20 qubits in presence of noise via density-matrix approaches demands an amount of memory equivalent to simulate 40 qubits with qHiPSTER. Even considering the necessity of a few thousands
noise realizations, our approach is more efficient for $N\gtrsim13$ (conservative estimate since $2^{13}=8192$).


\vspace{1cm}
\section{Scheduling quantum circuits on a square grid}
\label{app:scheduling}

Quantum circuits are usually described in terms of one- and two-qubit operations. The latter ones are defined by specifying which logical qubits (not to be confused with encoded qubits since
we do not consider active error correction techniques) they involve, without taking into account that such qubits may be associated with physical qubits that are placed in distant parts of the hardware.
To allow for a direct interaction, logical qubits must be routed to connected pair of physical qubits, and we use SWAP gates to achieve such task. We apply the framework presented in reference \cite{Guerreschi2018}
where the output of the compilation is presented in terms of the Physical Data Precedence Table (PDPT): each column corresponds to a physical qubit and each row to a clock-cycle (we assume that all gates
are executed in a single clock-cycle), the entry value is an integer that indicates if a qubit is idle (null entry) or if a gate from the algorithm is performed (positive entry, unique for each gate)
or if a SWAP gate for routing is performed (negative entry).

\begin{figure}[ht]
\vspace{5mm}
\includegraphics[scale=0.45]{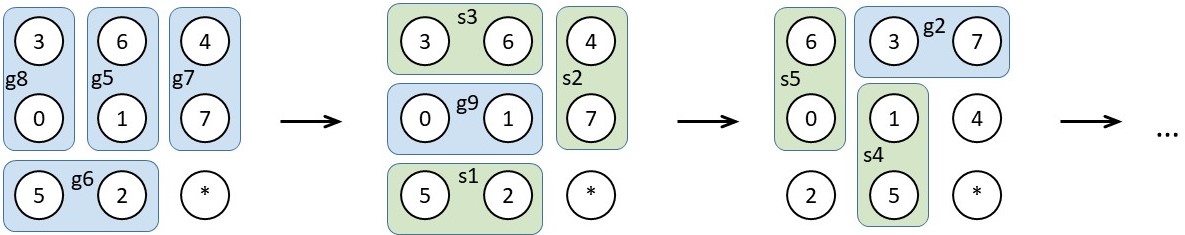}
\centering
\caption{First three snapshots of the schedule for the QAOA circuit. The logical instance involves 8 qubits and the schedule is compiled for a $3\times3$ grid of physical qubits,
here visualized as circles. Two-qubit gates from the QAOA circuits are depicted in blue, while the SWAP gates for routing are depicted in green. The number inside the circles
corresponds to the index of the logical qubit associated with the physical qubit. See text for additional details on the specific Max-Cut instance.}
\label{fig:schedule}
\end{figure}

Visualizing schedules for bi-dimensional architectures would require a tri-dimensional plot (the extra dimension being time), but one can present snapshot of the schedule for each clock-cycle separately.
In Fig.~\ref{fig:schedule} we provide an example of the first three clock-cycles of QAOA with 8 logical qubits on a machine that has $3\times3=9$ physical qubits. Notice that the first gate of the QAOA
algorithm are the two-qubit operations associated with the clauses of the Max-Cut instance, one gate per clause. For this specific instance, the edges of the graph are:
\begin{equation*}
    E=\big\{(7,6),(7,3),(5,3),(6,2),(6,1),(5,2),(7,4),(3,0),(1,0),(4,1),(5,4),(2,0) \big\}
\end{equation*}
and therefore the first gates of the QAOA circuits are:
\begin{eqnarray*}
  g_1 &=& \exp{(-i \gamma_1 Z_7 Z_6/2)} \\
  g_2 &=& \exp{(-i \gamma_1 Z_7 Z_3/2)} \\
  g_3 &=& \exp{(-i \gamma_1 Z_5 Z_3/2)} \\
  &\vdots& \\
  g_{12} &=& \exp{(-i \gamma_1 Z_2 Z_0/2)} \; .
\end{eqnarray*}

\newpage

The PDPT file is included for completeness (here the QAOA circuit has $p=4$).

\begin{center}
\begin{lstlisting}[basicstyle=\footnotesize,
caption={PDPT file with the qubit placement and complete schedule.
Each column corresponds to a physical qubit and each row to a clock-cycle: the entry value is an integer that indicates if a qubit is idle (null entry) or if a gate from the algorithm
is performed (positive entry, unique for each gate) or if a SWAP gate for routing is performed (negative entry).
The first three clock-cycles are visualized in Figure~\ref{fig:schedule}.},
captionpos=b,
frame=single]
# PDPT: each column is associated to a physical qubit , each row to a clock-cycle
## physical qubit indices ##################################################
        0       1       2       3       4       5       6       7       8
## logical qubit indices ###################################################
        3       6       4       0       1       7       5       2       *
############################################################################
        8       5       7       8       5       7       6       6       0
        -3      -3      -2      9       9       -2      -1      -1      0
        -5      2       2       -5      -4      0       0       -4      0
        0       0       0       4       11      11      4       0       0
        0       3       -8      -6      3       -8      -6      -7      -7
        12      16      0       12      18      -10     -9      -9      -10
        13      23      10      15      23      10      0       1       1
        28      28      17      26      26      14      0       19      20
        0       -11     -12     -13     -11     -12     -13     0       0
        -14     -14     0       0       -15     -15     24      24      0
        -16     29      29      -16     0       22      0       0       22
        -18     -18     0       31      31      36      0       -17     -17
        0       0       -19     38      27      -19     0       27      0
        -20     -21     -21     -20     0       25      0       0       25
        43      43      0       32      30      30      32      21      21
        0       0       0       33      37      34      35      40      39
        -22     0       0       -22     47      45      0       47      45
        48      48      0       46      50      50      46      41      41
        -25     -25     -23     51      51      -23     -24     -24     0
        0       49      49      58      57      0       0       44      44
        0       0       54      71      71      0       0       0       59
        0       -27     0       -28     -27     -26     -28     0       -26
        42      0       65      42      52      65      0       52      0
        56      70      70      60      53      0       0       55      0
        -29     0       0       -29     0       -30     0       0       -30
        67      67      0       68      68      0       0       64      64
        62      77      -31     62      0       -31     66      66      0
        -33     -33     0       63      69      69      63      -32     -32
        0       0       0       76      -34     74      78      -34     0
        0       61      0       0       61      0       0       72      72
        0       80      0       0       79      0       0       73      75
############################################################################
\end{lstlisting}
\end{center}


\section{Simulation of machines with additional qubits}
\label{app:ancilla-qubits}

It is realistic to assume that hardware devices may consist of more qubits than those required by the algorithm. Therefore it is important to understand the role of these extra qubits
and whether/how they should be included in the simulation.

Consider a quantum circuit involving $N$ qubits scheduled on a machine having $M>N$ physical qubits. For a lineal architecture an optimal schedule will not involve more than $N$ adjacent qubits,
and it is intuitive to neglect the remaining $M-N$ qubits when noise has no spatial correlations (i.e. acts on each physical qubit independently).

For a two-dimensional grid, the optimized schedule may involve more than $N$ qubits. While the actual state of the qubit register for the quantum circuit only span $N$ qubits, other qubits may be
involved in the routing procedure and, therefore, interact with ``logical'' qubits through SWAP operations. Notice that in this Section, as for the rest of the document, logical qubit does not refer
to an error corrected qubit, but to those qubits forming the logical register of the quantum algorithm.

Let us divide the physical qubit register in two parts: the logical register and the ancilla register. Their composition changes in time due to SWAP operations between qubits belonging to different registers.
The total state can be represented as $\rho_{L+A}$, the observable at the end of the circuit is $C_L$ (since it relates only to the logical register), and the quantity of interest is computed according to:
\begin{align*}
    \langle C \rangle &= \Tr_L \left[ \Tr_A \left[ \rho_{L+A} \right] C_L \right]\\
                      &= \Tr_{L+A}\left[ \rho_{L+A} C_L\otimes \mathbb{I}_A \right]\\
                      &= \frac{1}{R} \sum_{i=0}^{R-1} \bra{\psi_{L+A}^{(i)}} C_L\otimes \mathbb{I}_A \ket{\psi_{L+A}^{(i)}}  \, ,
\end{align*}
where the last line describe our method to introduce noise in the simulation by, effectively, representing the mixed state $\rho_{L+A}$ with a set of $R$ pure states
$\left\{ \ket{\psi_{L+A}^{(i)}} \right\}_{i=0,1,\dots,R-1}$. Each state $\ket{\psi_{L+A}^{(i)}}$ is generated by a perturbed version of the ideal quantum circuit in which many single-qubit,
stochastic gates are added. Since initially the logical and ancilla register are separable, this property is preserved until the end of each perturbed quantum circuit since SWAP gates cannot
create entanglement. Then $\ket{\psi_{L+A}^{(i)}}=\ket{\psi_L^{(i)}}\otimes\ket{\psi_A^{(i)}}$ and

\begin{align*}
    \langle C \rangle &= \frac{1}{R} \sum_i \bra{\psi_{L+A}^{(i)}} C_L\otimes \mathbb{I}_A \ket{\psi_{L+A}^{(i)}} \\
                      &= \frac{1}{R} \sum_i \bra{\psi_L^{(i)}} C_L \ket{\psi_L^{(i)}} \bra{\psi_A^{(i)}} \mathbb{I}_A \ket{\psi_A^{(i)}} \\
                      &= \frac{1}{R} \sum_i \bra{\psi_L^{(i)}} C_L \ket{\psi_L^{(i)}} \, ,
\end{align*}
meaning that the desired quantity can be computed by knowing $\ket{\psi_L^{(i)}}$ without specifying the state of the ancilla register. Therefore, the simulation only needs to include $N$ qubits
and not any of the extra qubits. Physically, this is due to the assumptions behind the noise model which does not include spatial or temporal correlations.


\section{Probability of observing the global solution}
\label{app:gs-population}

Recall the considerations expressed in the section named ``Comparison and performance crossover''. To access the quality of the solutions provided by QAOA, we report in FIG.~\ref{app:gs-population}
the overlap between the optimized state $\ket{\gamma,\beta}$ and the exact solution.
Notice that even achieving a probability of observing the solution assignment of only $0.1\%$ would imply that, with high probability, at least one of the 10,000 experimental repetitions of the
optimized circuit returns the global solution.

\begin{figure}[ht]
\vspace{-2mm}
\includegraphics[scale=0.4]{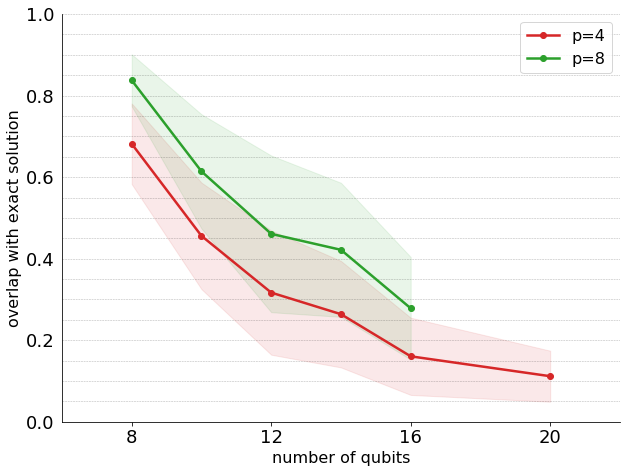}
\centering
\caption{Overlap with the global solution of the Max-Cut problem for the state at the end of the QAOA circuit after optimization, plotted as a function of the number of qubits or, equivalently, graph vertices. Each point represents the average over 40 instances and the shaded area corresponds to one standard deviation of the distribution.}
\label{fig:gs-population}
\end{figure}


\section{Statistics for QAOA results}
\label{app:QAOA-statistics}

The results presented in the insert of FIG.~2 (main text) 
have such small error bars that they are covered by the data markers. In this section, we report the error bar values and describe their significance. First of all, recall the meaning of each datapoint.
It represents the absolute time to solve with quantum computers instances of Max-Cut at a certain system size.

For each instance multiple optimization runs are performed, each starting from a different random value of the QAOA parameters, and only the one that achieves the best approximation ratio is associated
with the actual (approximate) solution of the instance. However, all runs are required to avoid being deceived by a local maximum. The cost of each optimization run must then be summed up.

Different instances may require a different cost to achieve their solution, even if the number of qubits is the same. This is due to the specific landscape in parameter space that may be more or less
smooth and/or present more or fewer local maxima. The time cost to solve a single instance is then the average cost over a set of instances, 40 in our study. From the statistical deviation of the
single-instance cost one can associate an error bar to the estimate of the \emph{average} cost according to
\begin{equation}
    \text{\small std. dev. of average}
        = \sqrt{\frac{\text{\small variance over instances}}{\text{\small number of instances}}} \, .
\end{equation}

Since the standard deviation associated to each point is smaller than the marker in FIG.~2 (main text)
, the values are provided in Table~\ref{table:instance-statistics} below.

\begin{table}[h!]
\centering
\begin{tabular}{ | c || c | c | c |}
\hline
    \hspace{4mm}$N$\hspace{4mm} & \hspace{3mm}$p=4$\hspace{3mm} & \hspace{3mm}$p=8$\hspace{3mm} \\
\hline \hline
     8 & 100.6 $\pm$ 0.7 & 211.3 $\pm$ 0.9 \\ \hline
    10 & 102.8 $\pm$ 0.7 & 220.6 $\pm$ 1.2 \\ \hline
    12 & 106.6 $\pm$ 0.6 & 226.2 $\pm$ 1.5 \\ \hline
    14 & 107.5 $\pm$ 0.9 & 233.2 $\pm$ 1.1 \\ \hline
    16 & 113.1 $\pm$ 1.0 & 241.1 $\pm$ 1.1 \\ \hline
    20 & 118.8 $\pm$ 1.0 &                 \\ \hline
\end{tabular}
\vspace{0.3cm}
\caption{Average cost of solving a single instance of Max-Cut with QAOA together with the corresponding standard deviation. The cost corresponds to the absolute time, in seconds, to run the quantum
algorithm on realistic hardware. Values correspond to the red and green datapoints in FIG.~2 (main text)
.}
\label{table:instance-statistics}
\end{table}


\vspace{1cm}
\section{AKMAXSAT running times}
\label{app:akmaxsat}

The classical solver used in this study is AKMAXSAT \cite{Kugel2012}, often utilized to benchmark Adiabatic Quantum Optimization devices. For the reduction of Max-Cut to Max-2-SAT we followed the same approach described in reference \cite{Gramm2003} and that requires a binary variable for each vertex of the graph to be partitioned and two clauses for each edge. Explicitly one has:
\begin{align*}
    \text{node of graphs: }x_1,x_2 \quad&\rightarrow
        \quad \text{binary variables: }x_1,x_2 \, ;\\
    \text{edge connecting }x_1\text{ and}x_2 \quad&\rightarrow
        \quad \text{clauses: }(x_1 \lor x_2)\, , \,(\lnot x_1 \lor \lnot x_2) \, .
\end{align*}
It is straightforward to see that at least one clause is always satisfied and that both clauses are satisfied if and only if the edge can be cut, \emph{i.e.} when $x_1=\lnot x_2$. Therefore,
if the graph for the Max-Cut problem has $E$ edges, the corresponding Max-SAT instance has $2 E$ clauses. If the maximum number of clauses that can be satisfied at the same time is $E+k$
(with $0<k\leq E$), the solution to the corresponding Max-Cut instance is $k$.

\begin{figure}[th!]
\includegraphics[scale=0.6]{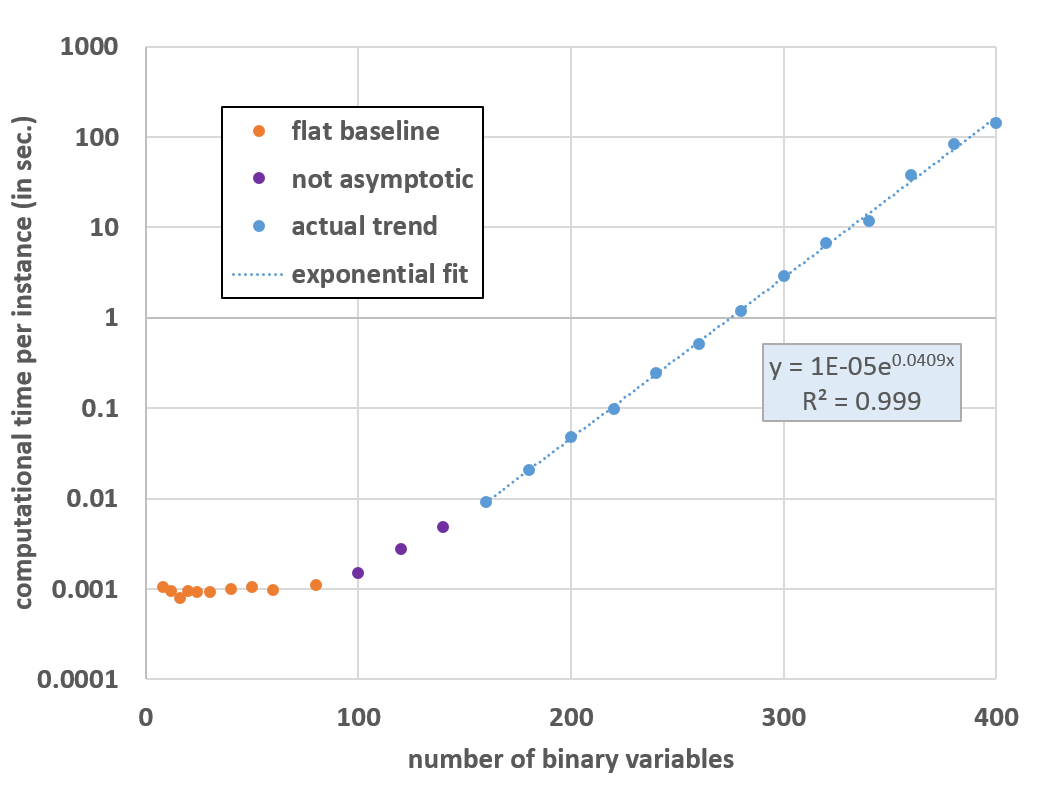}
\centering
\caption{Computational cost, in absolute time and averaged over 400 instances, to solve a single instance of Max-Cut as a function of the number of nodes in the graph,
corresponding to the number of logical qubits for the QAOA algorithm. The classical solver is AKMAXSAT and it was run on a single-socket Intel Xeon Phi 7250 (68 core Knights Landing per socket,
turbo enabled, flat, quad, 96 GB).}
\label{fig:classical-only}
\end{figure}

FIG.~\ref{fig:classical-only} provides the computational cost, in absolute time, required by AKMAXSAT to solve one instance (averaged over 400 instances). An exponential fit is provided,
together with the regression value $R^2=0.9963$. The timings have been obtained using the SSR-NFS computing cluster at PCL (Intel Labs). Jobs were run on a single Intel Xeon Phi 7250 node
(68 cores Knights Landing) in parallel, one per core
(Performance results are based on testing as of December 20th 2017 and may not reflect all publicly available security updates. See configuration disclosure for details. No product can be absolutely secure.
Intel, Xeon, and Intel Xeon Phi are trademarks of Intel Corporation in the United States and other countries. Other names and brands may be claimed as the property of others.).

\end{document}